%% file: main.tex
\documentclass{article}

\input{Preamble.tex}

\input{Commands.tex}

\title{TopROI: A topology-informed network approach for tissue partitioning}

\author{Sergio Serrano de Haro Iváñez\thanks{Mathematical Institute, University of Oxford, Oxford, UK.} \and Joshua W. Moore\footnotemark[1] \and Lucile Grzesiak\thanks{Centre for Human Genetics, University of Oxford, Oxford, UK} \and Eoghan J. Mullholand\footnotemark[2] \and Heather Harrington\footnotemark[1] $^,$\footnotemark[2] $^,$\thanks{Max Planck Institute of Molecular Cell Biology and Genetics, Dresden, Germany} $^,$\thanks{Center for Systems Biology Dresden, Dresden, Germany} $^,$\thanks{Faculty of Mathematics, TU Dresden, Dresden, Germany} \and Simon J. Leedham\footnotemark[2] \and Helen M. Byrne\footnotemark[1] $^,$\footnotemark[2] $^,$\thanks{Ludwig Institute of Cancer Research Oxford Branch, University of Oxford, Oxford, UK}}

\date{}

\makeatletter
\renewcommand{\@fnsymbol}[1]{\@arabic{#1}}
\makeatother

\begin{document}
\maketitle

\begin{abstract}
Mammalian tissue architecture is central to biological function, and its disruption is a hallmark of disease. Medical imaging techniques can generate large point cloud datasets that capture
changes in the cellular composition of such tissues with disease progression.
However, regions of interest (ROIs) are usually defined by quadrat-based methods that ignore intrinsic structure and risk fragmenting meaningful features. Here, we introduce TopROI, a topology-informed, network-based method for partitioning point clouds into ROIs that preserves both local geometry and higher-order architecture. TopROI integrates geometry-informed networks with persistent homology, combining cell neighbourhoods and multiscale cycles to guide community detection. Applied to synthetic point clouds that mimic glandular structure, TopROI outperforms quadrat-based and purely geometric partitions by maintaining biologically plausible ROI geometry and better preserving ground-truth structures. Applied to cellular point clouds obtained from human colorectal cancer biopsies, TopROI generates ROIs that preserve crypt-like structures and enable persistent homology analysis of individual regions. This study reveals a continuum of architectural changes from healthy mucosa to carcinoma, reflecting progressive disorganisation in tissue structure. TopROI thus provides a principled and flexible framework for defining biologically meaningful ROIs in large point clouds, enabling more accurate quantification of tissue organization and new insights into structural changes associated with disease progression.
\end{abstract}

\clearpage
\section{Introduction}

Cellular architecture and tissue organisation are critical determinants of biological function, and their disruption is a hallmark of disease. Morphological assessment of tissue geometry underpins key diagnostic criteria: for example, loss of crypt organisation correlates with colorectal cancer progression \cite{fleming2012colorectal}, glandular architecture forms the basis of the Gleason grading system in prostate cancer \cite{gleason1966classification, gleason1974prediction}, and changes in the cellular composition of 
red and white pulp 
informs lupus spleen pathology \cite{jacobson1995anatomy}.

Large-scale image analysis technologies (e.g. \texttt{QuPath} \cite{bankhead2017qupath}, \texttt{HALO} \cite{indicalabs_halo}) generate point cloud data that captures both the phenotypes and the positions of individual cells, offering unprecedented opportunities to study tissue architecture. These advances underpin the field of spatial biology, which seeks to map cellular organisation and relate it to biological function, disease progression, and response to treatment. Realising this potential requires addressing a key challenge: generated datasets are often too large and heterogeneous to analyse directly, and must first be  subdivided into smaller regions of interest (ROIs) for both biological interpretation and computational feasibility. The way in which ROIs are defined is crucial: they must preserve tissue architecture if downstream analyses are to reflect the underlying biology. Current practice typically partitions samples into quadrats of fixed size \cite{vipond2021multiparameter, smith2021developing, bull2024muspan}. Such grid-based ROIs neglect intrinsic tissue organisation and may fragment key structural features, potentially compromising subsequent analyses.

Advances in computational geometry and topology offer new opportunities to address this limitation of spatial biology. On the one hand, network science provides a mathematical framework for representing complex systems of many interacting units, enabling the study of structural patterns such as hierarchies \cite{clauset2008hierarchical}, communities \cite{porter2009communities, fortunato2016community}, and diffusion processes \cite{pastor2001epidemic}; in biomedical applications, methods from network science have been used to quantify cell phenotypes and interactions from mIHC data \cite{schapiro2017histocat}, analyse breast cancer cell neighbourhoods \cite{jackson2020single}, model large-scale brain connectivity \cite{he2010graph}, 
and conceptualise disease as perturbations of molecular interaction networks \cite{barabasi2011network}. We note the work of \cite{jackson2020single}, in which community detection was used to analyse geometric cell-cell contact networks. On the other hand, persistent homology \cite{otter2017roadmap} is a technique from topological data analysis that has found applications across a wide range of fields, serving as a tool to extract qualitative topological information in a way that is robust to noise and meaningful across scales. In spatial biology, it has been used to classify cell types in IgA nephropathic kidney \cite{benjamin2024multiscale}, characterise bone tissue microarchitecture \cite{pritchard2023persistent}, and detect structural changes in lupus-affected spleen \cite{torras2025topology}. Within oncology \cite{bukkuri2021applications}, it has been applied to analyse immune cell infiltration \cite{vipond2021multiparameter,bull2024integrating}, quantify architectural
changes during prostate cancer development \cite{lawson2019persistent}, and compare tumour vascular networks \cite{stolz2022multiscale}.

Here, we introduce TopROI, a framework that combines Delaunay geometric networks with features derived from persistent homology to partition point clouds into communities that preserve both local connectivity and higher-order architecture. Using synthetic data, we demonstrate that TopROI produces ROIs that are geometrically consistent and visually feasible, while accurately preserving ground-truth structural features.

We then apply TopROI to human colorectal tissue, a setting in which architecture plays a central role in cancer progression. As mutations disrupt cell function \cite{fouad2017revisiting}, intestinal crypts (tubular glands lined by a single layer of epithelial cells) progressively lose their circularity and thinness, and benign lesions evolve into malignant carcinomas. In 2D \textit{en face} tissue sections, crypts appear as empty cell circles, which become increasingly distorted and fragmented with disease progression. By defining ROIs that preserve such structures, TopROI enables finer-grained analysis of tissue heterogeneity, and reveals the architectural continuum from healthy mucosa through benign lesions to invasive cancer.

The remainder of this paper is structured as follows. We first introduce the algorithm for TopROI. The Data section then introduces the synthetic and clinical tissue datasets used for evaluation. In the Results, we present the performance of TopROI on synthetic data and demonstrate its application to human colorectal biopsies. In the Discussion, we examine the strengths and limitations of TopROI and outline potential directions for future research. Finally, in Materials and Methods we provide more detailed descriptions of both the datasets and the techniques used for their analysis.

\section{TopROI: topology-informed point cloud partitioning}

\begin{figure}[H]

    \centering
    \includegraphics[width=\textwidth]{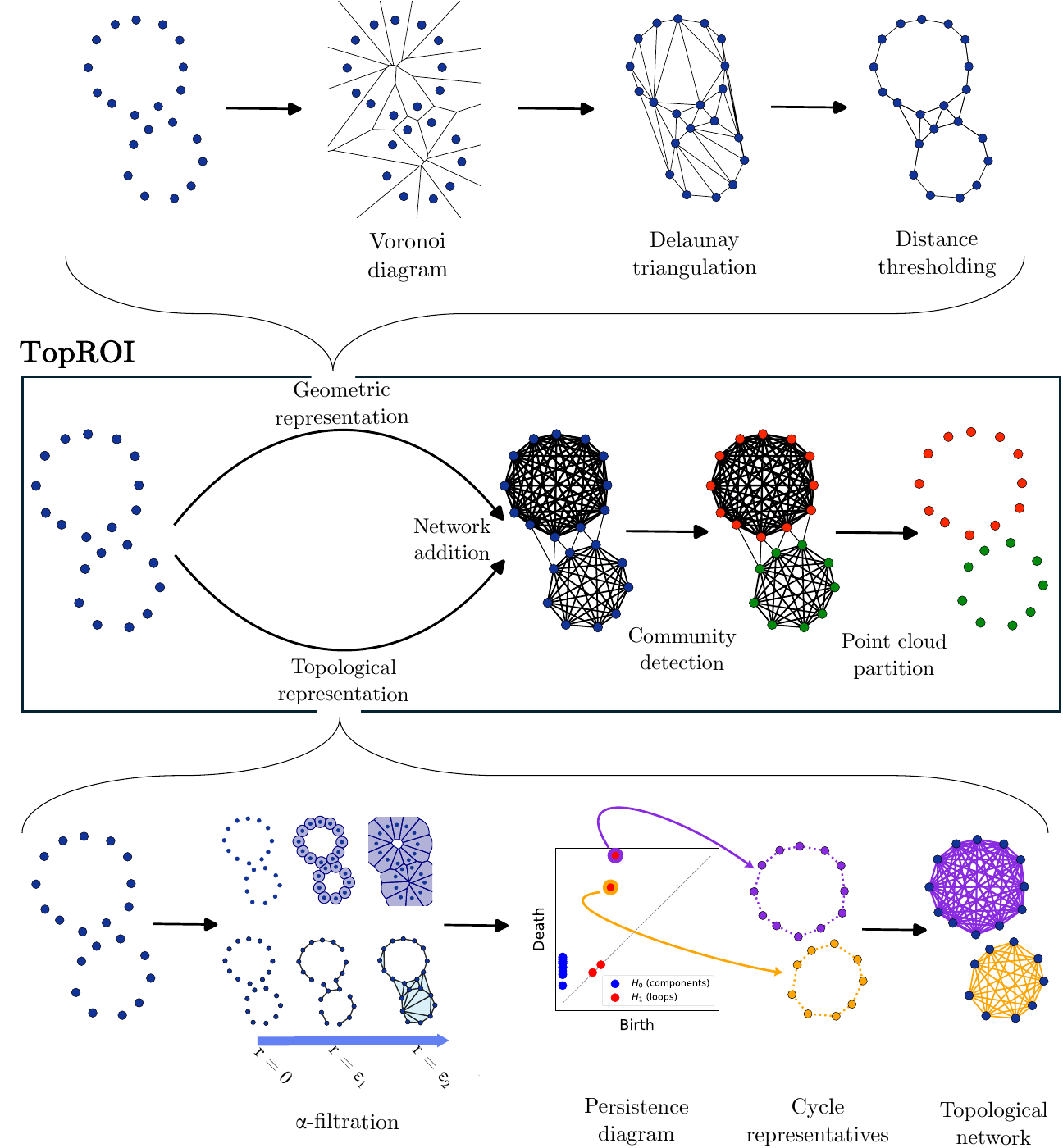} 
    \caption{Overview of TopROI pipeline. \textbf{Top:} construction of a Delaunay-based spatial graph, commonly used to represent cell contact networks; \textbf{centre:} full pipeline of TopROI; \textbf{bottom:} persistent homology extracts topological information from point clouds, in our case physical representatives of data features.}
    \label{fig:Explanation}
\end{figure}

Let $X \subset \mathbb{R}^2$ be a point cloud. TopROI constructs a weighted graph $G = (V, E, W)$, with vertices $V = X$, that integrates both geometric and topological information. The ROIs of $X$ are then defined as the detected communities of $G$. Figure \ref{fig:Explanation} illustrates the pipeline. 

We first construct the Delaunay triangulation $G^\text{geo} = (V, E^\text{geo})$, a geometric network representing local relationships (see Section \ref{subsec:methods_geometric_networks}). Edges longer than a threshold parameter $l > 0$ are removed from the network. For $u,v\in V$, the weight of the edge $(u,v)$ is
\[
w^\text{geo}(u,v) = \begin{cases}
1, & \text{if } (u,v)\in E^{\mathrm{geo}} \text{ and } \|u-v\|\leq l \\
0, & \text{otherwise}
\end{cases}
\]
where edges of weight 0 are not present in the network.

On the other hand, from the $\alpha$-complex filtration of $X$, we compute 1D persistent homology class representatives (see Section \ref{subsec:methods_cycle_networks}), physical realisations of the topological features of the point cloud.  Optionally, representatives may be filtered out according to their persistence, birth, or death values, to ensure that the resulting network encodes features within the scale of interest.  We then identify each 1D representative with the nodes incident to its edges, $C \subset V$, and insert a complete subgraph on $C$ with weights
\[
w^{top}(u,v) = \max_{C':\ u,v\in C' } p(C'),
\]
where $C'$ ranges across representatives and $p(C')$ denotes their persistence. Edges whose nodes do not share any representative receive weight 0. This yields the topological network $G^\text{top} = (V, E^\text{top}, W^\text{top})$.

The final network $G = (V, E, W)$ is defined as the weighted union
\[
E = E^\text{geo}\ \cup\ E^\text{top}, \quad w(u,v) = \max\{w^\text{geo}(u,v), w^\text{top}(u,v)\}.
\]
Communities of $G$ are obtained by partitioning the network via the Leiden algorithm, optimising the Reichardt and Bornholdt's Potts model quality function with resolution parameter $\gamma$ (see Section \ref{subsec:methods_community_detection}). The value of $\gamma$ regulates the resulting number of communities, and is chosen via a secant search (see Appendix \ref{App:Res_param_det}) to produce the desired number. Resulting communities are the proposed geometrically and topologically faithful ROIs of $X$.

\section{Data}
We analyse two datasets: one synthetic and one derived from human colorectal tissue (see Section \ref{subsec:methods_data} for detailed descriptions).

\textbf{Synthetic dataset.} To evaluate TopROI under controlled conditions, we generated synthetic point clouds designed to mimic glandular tissue organisation ($n=30$). These datasets included gland-like circular and elliptical structures embedded in a noisy cellular background, with perimeter cells serving as references for ellipse preservation.

\textbf{Pathology-graded colorectal dataset.} We also analysed $n=15$ human colorectal resection biopsy samples, stained by immunohistochemistry and imaged with the \texttt{HALO} platform \cite{indicalabs_halo}. Each sample was represented as a 2D point cloud of cell centroids, with cell types identified but not used in the present study. Samples were annotated by trained pathologists according to morphological features observed in H\&E-stained images, and were histologically classified into three groups: healthy mucosa ($n=3$), benign precursor lesions or polyps ($n=6$), and cancers ($n=6$).

\section{Results}

\subsection{TopROI preserves topology and geometry in synthetic point clouds}

\begin{figure}[h!]

    \centering
    \includegraphics[width=\textwidth]{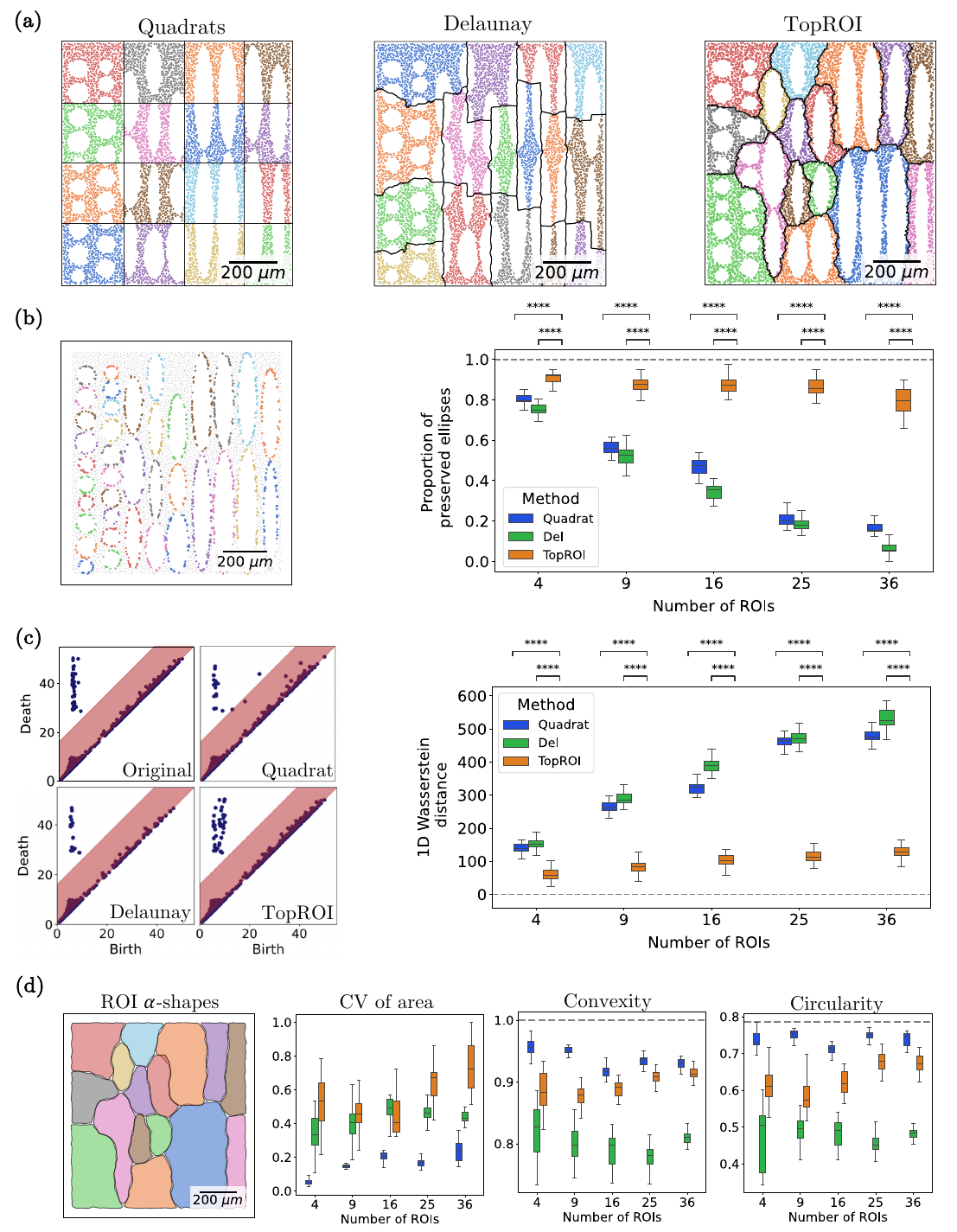} 
    \caption{TopROI preserves geometry and topology in synthetic data. \textbf{(a)} Partition of synthetic dataset into 16 ROIs (black Voronoi borders are not a result of the algorithm, and are included for visualisation only); \textbf{(b)} ellipse markers example, and ellipse preservation results;  \textbf{(c)} comparison of original and reconstructed 1D persistence diagrams with noise band (red), and average Wasserstein distances;   \textbf{(d)} Example alpha shapes of TopROI ROIs, and average area variation, circularity ($\nicefrac{\pi}{4}$ for a square, dashed) and convexity (1 for a square, dashed).}
    \label{fig:Benchmark}
\end{figure}

We first evaluated TopROI on the synthetic dataset (see Section \ref{subsec:methods_evaluation_synthetic} for details). Three methods were used to partition each point cloud 
into ROIs: grid-based (quadrats), community detection on the truncated Delaunay network (i.e., the geometric cell contact network, \cite{jackson2020single, bull2024muspan}), and TopROI. The number of ROIs generated was matched between methods (Figure \ref{fig:Benchmark}a).

TopROI substantially outperformed alternative methods in preserving ground-truth elliptical structures, maintaining, on average, over 85\% of ellipses at low ROI numbers, and declining only modestly at finer resolutions (Figure \ref{fig:Benchmark}b). It also achieved markedly better agreement between the original and reconstructed persistence diagrams, with Wasserstein distances 2–4 times lower than those generated from quadrats or Delaunay partitions (Figure \ref{fig:Benchmark}c). Finally, the TopROI regions preserved high convexity and circularity and exhibited greater size variation, reflecting the method’s ability to generate geometrically reasonable partitions while adapting the ROIs to local structural properties (Figure \ref{fig:Benchmark}d). Together, these results demonstrate that TopROI preserves both the topology and geometry of the synthetic dataset more effectively than existing approaches.

\clearpage

\subsection{Quantifying architectural transition and heterogeneity in colorectal cancer progression}\label{subsec:Results_real}

\begin{figure}[h!]
    \centering
    \includegraphics[width=\textwidth]{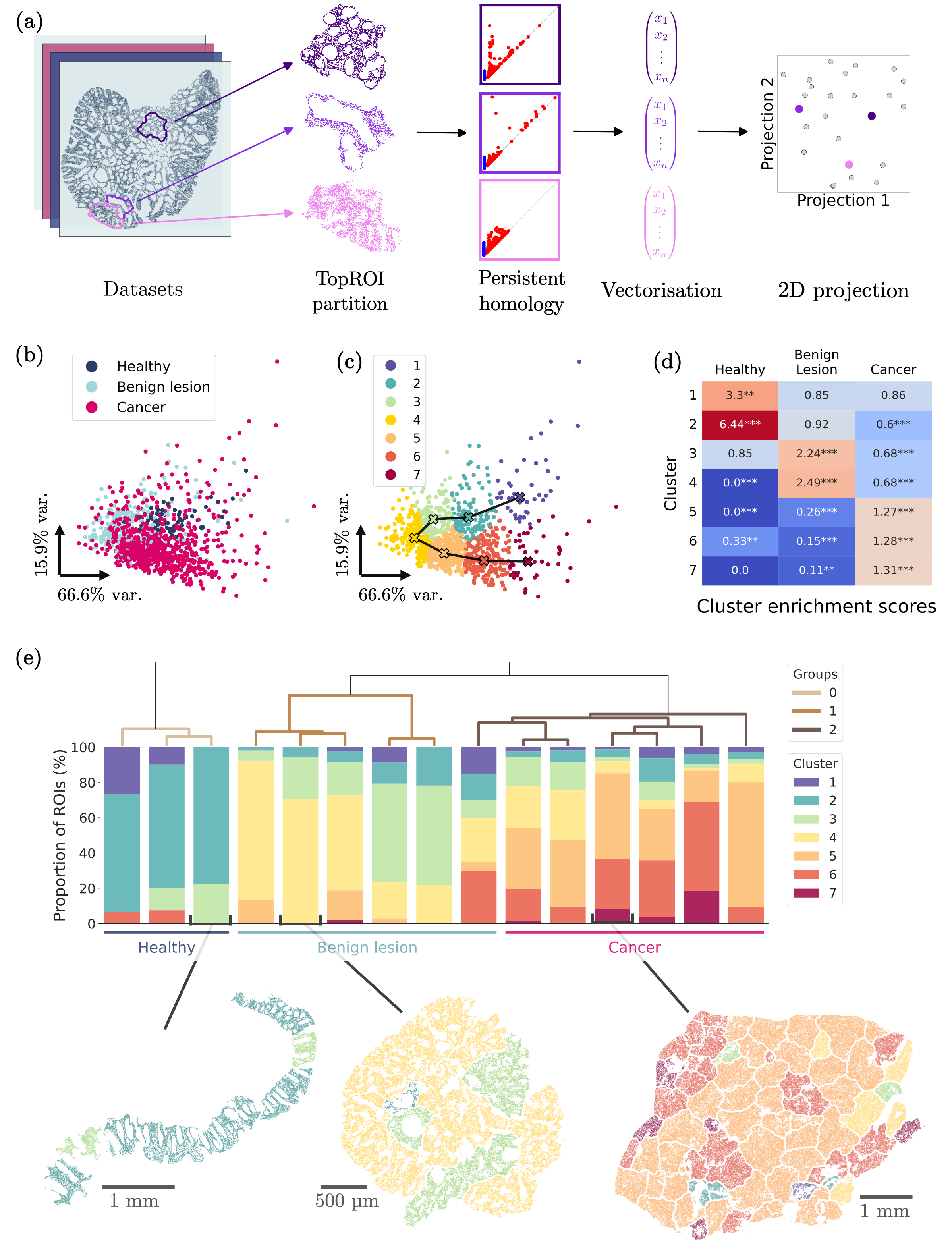} 
    \caption{Topological analysis and clustering of ROIs from biopsied tissue samples. \textbf{(a)} Topological analysis pipeline. \textbf{(b)} 2D PCA projection of vectorised features. \textbf{(c)} K-means clustering results projected onto PCA space, and minimum spanning tree connecting cluster centroids (black line); cluster IDs follow tree order. \textbf{(d)} Disease stage enrichment scores for each topological cluster. \textbf{(e)} Topological profiles of individual samples, grouped via hierarchical clustering (\textit{Groups}, brown shades). Groups are obtained with the default parameters of \texttt{scipy.cluster.hierarchy} \cite{2020SciPy-NMeth}.}
    \label{fig:Real_PCA}
\end{figure}

We also applied TopROI to human colorectal biopsy samples histologically classified as healthy mucosa, benign lesion, or carcinoma. Each sample was partitioned into ROIs of average size $\sim 250,000 \mu m^2$, yielding a total of 64 ROIs from healthy tissue samples, 218 from benign lesions, and 826 from cancers. For each ROI, we computed persistent homology, vectorised the resulting persistence diagrams into topological descriptors, and analysed the resulting feature vectors to compare tissue architecture across disease states (Figure \ref{fig:Real_PCA}a, see Section \ref{subsec:methods_evaluation_real} for details).

Analysis of ROI topologies revealed clear differences across disease stages. Principal component analysis (PCA) of vectorised persistence diagrams (Figure \ref{fig:Real_PCA}b) showed that 82.5\% of variation in ROI topology was captured by the first two principal components. K-means clustering of persistence vectors grouped ROIs into seven clusters (see Section \ref{subsec:methods_evaluation_real} for elbow test details), with the minimum spanning tree of cluster centroids (Figure \ref{fig:Real_PCA}c) reflecting disease progression: clusters 1–2 were enriched for healthy ROIs, clusters 3–4 for benign lesions, and clusters 5–7 for cancers (Figure \ref{fig:Real_PCA}d).

When examining individual datasets, ROIs originated from the same sample were often assigned to different topological clusters. For each sample, we measured the proportion of ROIs belonging to each cluster, generating a topological profile. Hierarchical clustering of these topological profiles classified samples into three major groups that corresponded closely to the pathologist-defined stages of healthy, benign, and cancerous tissue (Figure \ref{fig:Real_PCA}e). These results demonstrate that TopROI partitions tissue into ROIs that preserve biologically relevant topological structure, capturing spatial patterns that evolve systematically from healthy to cancerous tissue.

\clearpage

\section{Discussion}

We have presented TopROI, a topology-enhanced geometric network approach for partitioning large point clouds into regions of interest that respect point-cloud architecture. Our results in synthetic data demonstrate that TopROI overcomes key shortcomings of conventional ROI strategies such as quadrats, which impose artificial boundaries that can fragment structures and affect downstream analyses of tissue architecture, and local geometric networks, which fail to capture higher-order features. By integrating persistent homology with Delaunay-based networks, TopROI yields partitions that simultaneously preserve local geometry and key architectural motifs such as glandular structures.

A central advantage of TopROI is its ability to adapt to resolution. By tuning the resolution parameter $\gamma$, the method can create ROIs and identify features across multiple spatial scales, from local architecture to broader tissue organisation. This adaptiveness is evident in the analyses of the synthetic dataset, where topological preservation remained robust across ROI sizes, and the human colorectal cancer dataset, where ROI partitions capture heterogeneity at scales relevant to colorectal cancer pathology. The approach is further strengthened by its modularity: networks can be tailored to emphasise different biological or geometric properties through simple adjustments such as edge or persistence thresholds, and weight assignments (e.g. weighting Delaunay edges to emphasise local geometry). The pipeline can be further customised to include other types of networks that reflect different point cloud characteristics; for datasets with cell type annotations, for example, one could construct networks that preferentially connect cell types known to share signaling or communication pathways, enabling preservation of biologically meaningful microenvironments. 

The use of persistent homology in TopROI relates to the biological questions of interest. Beyond ROI definition, our results highlight opportunities for quantitative analysis of tissue organisation in disease: in the colorectal cancer analysis, we showed that persistent homology captures the gradual loss of tissue architecture that characterises disease progression. TopROI allowed us to decompose 2D biopsies into ROIs whose topological descriptors, when vectorised and clustered, revealed a continuum from healthy mucosa to benign lesions to carcinoma. It evidenced the existence of both intra- and inter-sample morphological heterogeneity, while still being able to delineate between pathology stages. Altogether, this work illustrates the potential of persistent homology to provide quantitative insights into tissue organisation, heterogeneity, and its disruption in disease. While topological data analysis has previously been used to assess global loss of structure, our partition-based approach emphasises sample heterogeneity underlying disease progression. Although demonstrated in colorectal cancer, the same principles apply to other structurally organised tissues (such as the prostate, kidney, or spleen) where pathology involves region-specific alterations in organisation.


Nonetheless, several limitations remain. First, the method requires data-informed choices: thresholds for geometric networks edge length and topological features persistence, as well as calibration of the resolution parameter (here via a secant method), all influence ROI size and composition. While this flexibility is a strength, it also necessitates domain knowledge to ensure structurally meaningful outcomes. Secondly, the scalability of TDA restricts applications to very large point clouds. While $\alpha$-complexes are more efficient than alternative Vietoris–Rips or \v{C}ech complexes, determining cycle representatives with current implementations remains slow, making whole-tissue analyses challenging. Finally, selecting cycle representatives that accurately reflect geometric and visual boundaries is itself an open problem, and existing implementations may not consistently produce representations of the desired quality. In particular, not all structural features are equally well captured: the persistence of an elliptical cycle depends exclusively on its minor axis, potentially reducing its weight in the partitioning process. 

Future improvements could address these challenges. First, richer datasets, such as biopsy imaging with cell boundary markers or full cell shapes, would reduce the need for heuristic thresholds and approximations (e.g. inferring cell-cell contacts via Delaunay triangulations). Secondly, faster algorithms for computing cycle representatives would improve scalability and yield more faithful structural features. This could in particular enable the extension of TopROI to higher-dimensional data, a step currently limited by both computational costs and the difficulty of interpreting higher-order cycles. Finally, alternative formulations of persistent homology, such as ellipsoid-based approaches \cite{kalivsnik2024finding, kalivsnik2024persistent}, may better capture elongated or anisotropic structures that are underrepresented by classical methods. These advances would broaden applicability of TopROI and enhance its ability to detect meaningful architectural features.

In conclusion, TopROI provides a flexible, topology-informed framework for partitioning point clouds that preserves both geometry and higher-order architecture. By bridging computational geometry, network science, and topological data analysis, it offers a principled way to define biologically meaningful ROIs, enabling more faithful quantification of tissue structure. While computational and methodological challenges remain, the modularity of the approach positions TopROI as a general platform for tissue analysis across diverse contexts, with particular promise in applications where architectural disruption is a defining feature of disease.

\section{Materials and Methods}

\subsection{Mathematical background for TopROI}

\subsubsection{Geometric networks from spatial data}\label{subsec:methods_geometric_networks}
A network $G = (V,E)$ is defined by sets of nodes $V=\{v_i\}_{i\in I}$ representing individuals (e.g. cells) and edges $E = \{(v_{j_1}, v_{j_2})\}_{j_k \in I} \subseteq V \times V$ representing pairwise relationships (e.g. physical contact or cell signalling). Edges may have associated weights $W = \{ w(e)\}_{e\in E}$, scalar values that quantify interaction strengths. One classical method for constructing networks from spatial data is proximity graphs \cite{barthelemy2022spatial}, in which edges represent spatial relations between nodes embedded in a metric space.

We use the Delaunay triangulation, a proximity graph that is dual to the Voronoi diagram of a point cloud \cite{barthelemy2022spatial}. Given a discrete subset $X \subset \mathbb{R}^2$, the Voronoi region of $x \in X$ is 
\[
R_x = \{p \in \mathbb{R}^2\ :\ d(x,p) \leq d(x',p), \ \ \ x' \in X\},
\]
where $d$ is the Euclidean distance. Here, $R_x$ is the set of points that are closer to $x$ than to any other points $x' \in X$. The Delaunay triangulation of $X$ is a network with nodes $V=X$ and edges
\[
E = \{(x, x') :  R_x \cap R_{x'} \not= \emptyset \ \ \ \forall x,x' \in X\},
\]
so that two points in $X$ are connected if their Voronoi regions intersect (see Figure \ref{fig:Explanation} for an example). In biological applications, Delaunay networks can be used to estimate cell-cell contact networks \cite{dries2021giotto, edelsbrunner2004biological, 
bull2024muspan}, as Voronoi diagrams can be used to approximate the shape of cell boundaries from cell centroids \cite{
atienza2021stable, benjamin2024multiscale}. For improved biological realism, edges exceeding a specified length threshold (e.g. the maximum cell diameter) may be excluded to prevent biologically unrealistic interactions. We construct the Delaunay network using the \texttt{NetworkX} library \cite{hagberg2008exploring} implementation.

\subsubsection{Cycle networks from spatial data}\label{subsec:methods_cycle_networks}
While geometric networks capture local architecture, they do not account for higher-order structures that may be biologically relevant (e.g. the circular, empty border of a colorectal crypt). To incorporate these features into a network framework, we employ persistent homology \cite{otter2017roadmap}, a prominent tool from topological data analysis.

Point clouds themselves do not carry intrinsic topology. A standard approach to extract topological information from them is to replace each point by a ball of growing radius $r$, and then to study the emergence and disappearance of features as $r$ increases. In order to formalise this process, intersecting balls are endcoded by simplices: the original points correspond to 0-simplices, added at $r=0$; an edge (1-simplex) between two points is added  if their balls first intersect at radius $r = \varepsilon$; similarly, a triangular face (2-simplex) is added when three balls intersect, and so on. For computational efficiency, one can first intersect each ball with the Voronoi region of its corresponding point, thereby reducing distance computations -- this construction is known as the $\alpha$-complex \cite{otter2017roadmap} (see Figure \ref{fig:Explanation}).

The set of all simplices that exist at a fixed radius $r=\varepsilon$ forms a simplicial complex $K_\varepsilon$. Denoting by $C_n(K_\varepsilon)$ the vector space generated by $\mathbb{Z}_2$-linear combinations of the $n$-simplices of $K_\varepsilon$, one defines the boundary map $\partial^\varepsilon_n:C_n(K_\varepsilon) \rightarrow C_{n-1}(K_\varepsilon)$, which assigns to each $n$-simplex the sum of its oriented $(n-1)$-faces (or border). For example, the border of an edge consists of its two endpoints. The $n$th homology group of $K_\varepsilon$ is then
\[
H_n(K_\varepsilon) := \faktor{\ker \partial^\varepsilon_n\ }{\text{im } \partial^\varepsilon_{n+1}},
\]
whose elements correspond to $n$-dimensional holes: for $n = 0$ these are connected components, for $n = 1$, empty cycles, and so on.

As $r$ increases, one has an inclusion (or filtration) of complexes $K_0 \subset K_{\varepsilon_1} \subset K_{\varepsilon_2} \subset \cdots$, which induces a series of maps between homology groups, $H_n(K_0) \rightarrow H_n(K_{\varepsilon_1}) \rightarrow H_n(K_{\varepsilon_2}) \rightarrow \cdots$. This framework defines persistent homology, which tracks the evolution of the elements of the homology groups as the radius increases: for example, the radius at which they first appear (birth) or later vanish (death). The multiscale signature is visualised in a persistence diagram (Figure \ref{fig:Explanation}) a multiset $B = \{(\text{b}_i, \text{d}_i) \}_{i\in I}$ where each element is a pair of $(\text{birth}, \text{death})$ values that represents a topological feature. The persistence of a feature, death $-$ birth, quantifies the range of scales for which the feature is detectable. 

In applications, one may be interested in finding explicit representatives of persistent homology classes in the underlying point cloud (i.e., to identify subsets of points that give rise to specific topological features of the barcode). Since homology groups are equivalence classes, they have no canonical representative, and finding meaningful, computationally tractable representatives is an active area of research \cite{barbensi2022hypergraphs, vcufar2020ripserer, henselman2016matroid, li2021minimal, Bauer2021Ripser, maria2014gudhi}. 
Once representatives are identified, they can be used to construct a network that encodes the topological structure of the point cloud \cite{barbensi2022hypergraphs}: nodes correspond to points, and a weighted edge is placed between two nodes whenever they co-occur in at least one representative (see Figure \ref{fig:Explanation}). The edge weight is given by the maximum persistence among the cycles shared by the two nodes.

We use the \texttt{Ripserer} Julia library \cite{vcufar2020ripserer} to create the $\alpha$-filtrations and compute cycle representatives. For each cycle, we compute its shortest possible representative at the filtration time corresponding to $1/8$th of its lifespan. This ensures that the resulting cycle representatives are consistent with the visual geometric borders of the holes in the point cloud (see Appendix \ref{App:Cycle_lifespan}).

\subsubsection{Community detection methods}\label{subsec:methods_community_detection}
A central concept in network science are communities -- groups of nodes whose internal connections are denser than those with the rest of the network \cite{porter2009communities, fortunato2016community}. Community detection is a classical problem in network science, as community structure often reveals organisational properties of the underlying system (e.g. cells engaged in the same biological process). 


One way to identify disjoint communities in a network is the Leiden algorithm, \cite{traag2019louvain}, a refinement of the widely used Louvain algorithm \cite{blondel2008fast}. Both algorithms identify communities by finding a network partition that maximises a specified quality function. Given a network with nodes $V$, we use the Reichardt and Bornholdt’s Potts model (RBP) function \cite{reichardt2006statistical} to score a community partition $\{\mathcal{C}_v\}_{v\in V}$:
\[
Q = \sum_{u,v\in V} \left( A_{uv} -\gamma \frac{k_u k_v}{2m}\right)\ \delta(\mathcal{C}_u, \mathcal{C}_v),
\]
where $\gamma$ is a resolution parameter, $A_{uv}$ the edge weight between nodes $u$ and $v$, $m$ the sum of all edge weights, $k_v$ the weighted degree of node $v$, and $\delta(\mathcal{C}_u, \mathcal{C}_v)$ indicates whether nodes $u$ and $v$ belong to the same community ($=1$) or not ($=0$).

The RBP score quantifies the quality of a community partition by comparing observed connectivity within communities ($A_{uv}$ term) to the expected connectivity under a random null model that preserves node weighted degrees ($\nicefrac{k_u k_v}{2m}$ term). Higher values of $Q$ indicate more cohesive communities, with fewer edges between them. The parameter $\gamma$ controls resolution: for $\gamma =0$, any type of connection increases $Q$, and thus the optimal communities are the connected components of the network. The higher the value of $\gamma$, the more strongly interconnected a potential community needs to be to increase $Q$ -- and thus, smaller and better-connected communities are favoured.

Community detection is carried with the \texttt{leidenalg} Python library \cite{traag2023leidenalg} implementation.

\subsection{Data}\label{subsec:methods_data}
\textbf{Synthetic dataset.} To evaluate TopROI under controlled conditions, we generated a synthetic point cloud dataset that resembles glandular tissue organisation ($n=30$ samples). Each synthetic dataset consists of a region of approximately $\sim 950 \times 1000 \mu m^2$, where points are arranged in a triangular lattice with $10 \mu m$ side length, approximating typical bowel tissue cell size and spacing (see Table \ref{table:RealPointCloudData}).
To replicate glands cut at varying angles, we embedded 20–30 circles (radius 30–50$\mu m$), 10–20 ellipses (major axis 100–150$\mu m$, minor axis 25–50$\mu m$), and 6–10 ellipses (major axis 200–300$\mu m$, minor axis 25–50 $\mu m$). Points of the underlying triangular lattice that fell within these shapes were removed, creating gland-like luminal spaces. Along the perimeter of each ellipse, cells were uniformly distributed at a density of $0.08 - 0.15\ \nicefrac{\text{cells}}{\mu m}$, representing gland-forming cells. The perimeter cells were also used as a reference to assess crypt preservation. Gaussian noise with variance $\sigma = 2\mu m$ was applied to all points to simulate measurement variability. For the purpose of our study all points are assumed to be identical, with no cell type assignment.

\textbf{Pathology-graded colorectal cancer dataset.} We analysed $n=15$ human colorectal cancer biopsy samples processed by immunohistochemistry staining and imaged with the \texttt{HALO} platform \cite{indicalabs_halo}. The resulting data consisted of 2D point clouds, where each point corresponds to the $(x,y)$ centroid of a cell. Cell coordinates were obtained through nuclear segmentation and image analysis conducted using \texttt{HALO}. An immune panel stain (CD4, CD8, FOXP3, ECAD, CD68, MPO, DAPI) enabled identification of seven cell types: epithelium, regulatory T cells, cytotoxic T cells, T helper cells, macrophages, neutrophils, and unclassified cells -- although in the present work, we do not distinguish different cell types. For more details of mIHC immune population exploration, see \cite{vazquez2022dynamic, koppens2021bone}. The point clouds were curated to remove artefactual points located at tissue borders or within crypts if they were located more than $30 \mu m$ from their nearest neighbour, a distance considered biologically implausible and likely arising from tissue processing artifacts or erroneous cell detection. Samples were histologically classified by a trained pathologist into three categories: healthy mucosa ($n=3$), benign lesions ($n=6$), and cancerous lesions ($n=6$). See Table \ref{table:RealPointCloudData} for a summary of the characteristics of the data, and Appendix \ref{App:Real_datasets} for plots of the point clouds corresponding to all samples.

\begin{table}[h]
\centering
\begin{tabular}{l|c|c|c|}
\cline{2-4}
& \multicolumn{1}{c|}{\textbf{Normal mucosa}}  & \multicolumn{1}{c|}{\textbf{Benign lesion}} & \multicolumn{1}{c|}{\textbf{Cancer}}\\ \hline
\multicolumn{1}{|l|}{Samples}                    & 3                                            & 6
& 6 \\ \hline
\multicolumn{1}{|l|}{Total cells}                & $1.5\cdot 10^4$ -- $5.9\cdot 10^4$                                 & $3.7\cdot 10^4$ -- $1.4\cdot 10^5$ & $1.6\cdot 10^5$ -- $3.8\cdot 10^5$                               \\ \hline
\multicolumn{1}{|l|}{\makecell{Bounding box \\ ($mm \times mm$)}}                  & \makecell{$2.9 \times3.7 $  --\\$10.0 \times 4.7$} & \makecell{$2.5\times 2.6$ -- \\ $5.2\times 5.2$} & \makecell{$6.0\times 4.4$ -- \\ $7.8\times 17.1$}  \\ \hline
\multicolumn{1}{|l|}{\makecell{Approximate \\ area ($m m^2$) }}                  & $2.4$ -- $10.0$  & $4.3$ -- $13.3$ & $19.1$ -- $40.2$  \\ \hline
\multicolumn{1}{|l|}{\makecell{Average distance \\ of 6 nearest \\ neighbours ($\mu m$)}}                   & $11.8 \pm 2.5$ -- $12.6 \pm 2.9$ & $10.2 \pm 1.7$ -- $11.2 \pm 2.7$ & $10.7 \pm 0.8$ -- $11.9 \pm 2.2$  \\ \hline
\end{tabular}
\caption{Summary of the point clouds obtained from the human colorectal cancer tissue biopsies. Values represent the range of each characteristic. We note that in 2D, the average degree of a Delaunay graph approaches 6 as the number of points increases.}\label{table:RealPointCloudData}
\end{table}

\subsection{Evaulation of synthetic datasets}\label{subsec:methods_evaluation_synthetic}

To assess TopROI, we used synthetic point clouds designed to resemble cellular structure with glandular arrangements. In the Delaunay network, representing cell–cell contacts and thus tissue geometry, we removed all edges longer than $20\ \mu m$ (twice the simulated cellular diameter, to account for noise in cell position). For the topological network, we excluded features with persistence $< 5\ \mu m$ (corresponding to the simulated cell radius, below which structural gaps are indistinguishable from variations in cell separation). No upper persistence threshold was applied, as the synthetic data were designed without large-scale artefactual features.

Three methods were used to partition each dataset into ROIs: regular grid (quadrats), community detection on the truncated Delaunay network, and TopROI. The number of ROIs was matched across methods (4, 9, 16, 25, 36), corresponding to quadrat side lengths of 500–167 $\mu m$. For graph-based methods, the resolution parameter $\gamma$ was tuned using a secant search (see Appendix \ref{App:Res_param_det}) to achieve the target number of communities.

Topological preservation was assessed using two metrics. First, ellipse preservation was quantified by requiring at least 90\% of the points on the perimeter of each ellipse to lie within the same ROI. Secondly, we compared 1D persistence diagrams of the full point cloud with those reconstructed from partitions: for a specific partition, persistence diagrams were computed for each ROI and aggregated. Similarity between the original and reconstructed diagrams was measured with the 1-Wasserstein distance: given two persistence diagrams $B=\{\ (b_i,d_i)\ \}_{i\in I}$, $B'=\{\ (b'_j,d'_j)\ \}_{i\in J}$, their 1-Wasserstein distance is
\[
    W_1(B, B') := \inf_{\phi:\ B \rightarrow B'} \sum_{(b, d) \in B} || (b, d) - \phi\left( (b,d ) \right) ||_2,
\]
where $||\cdot||_2$ is the Euclidean distance, and the infimum is taken over all bijections $\phi$ between $B$ and  $B'$. To ensure such bijections exist, both diagrams are augmented with points of the form $(x,x)$ (the diagonal), each with infinite multiplicity. To prevent low-persistence noise from dominating the metric, we applied the bootstrap method (n=1000 samples) of \cite{fasy2014confidence} to estimate 95\% confidence noise bands, which were ignored for the Wasserstein distance. To reduce computation times, we used the sliced 1-Wasserstein distance (\cite{maria2014gudhi}, 100 directions).

Geometric properties of ROIs were quantified using the \texttt{MuSpAn} Python package \cite{bull2024muspan}. Alpha shapes ($\alpha=1/50$) were generated for all ROIs, and consistency in ROI size and shape was evaluated using the coefficients of variation of ROI area, convexity ($\nicefrac{\text{area}}{\text{convex hull area}}$), and circularity ($\nicefrac{4\pi\text{ area}}{\text{perimeter}^2}$).

\subsection{Evaluation of colorectal cancer dataset}\label{subsec:methods_evaluation_real}
TopROI was applied to human colorectal biopsy samples classified by pathologists as healthy mucosa, benign lesions, or cancer. In the Delaunay network, edges longer than $30\ \mu m$ were removed. For the topological network, thresholds were chosen to ensure that TopROI captured biologically relevant structures such as intestinal crypts. Specifically, cycle representatives with persistence $< 5\ \mu m$ (subcellular gaps) or $> 100\ \mu m$ (artefactual features, such as tears introduced during tissue preparation; see Appendix \ref{App:Artefactual_figures}) were excluded. Features with birth $> 25\ \mu m$ were also filtered out, since biological crypts are delineated by epithelial cells and therefore emerged earlier in the filtration.

Each sample was partitioned into ROIs of average area $250\ 000 \mu m^2$ (equivalent to 500 × 500 µm$^2$ quadrats). Resolution parameters for community detection were determined using a secant method to produce the desired number of ROIs (see Appendix \ref{App:Res_param_det}). Across all samples, this resulted in 64 ROIs generated from partitioning the 3 healthy mucosa datasets, 218 from the 6 benign lesions, and 826 from the 6 carcinomas.

Persistent homology of each ROI was computed using the $\alpha$-complex filtration implementation of \texttt{Ripserer} \cite{vcufar2020ripserer}. Persistence diagrams were then vectorised following the approach of \cite{ali2023survey}, implemented in \texttt{MuSpAn} \cite{bull2024muspan}, with certain statistics excluded: first, all statistics related to the birth of $H_0$ features (mean, standard deviation, and 10/25/50/75/90\% quartiles) were omitted, since by construction all $H_0$ features are born at time $t=0$. In addition, the counts of $H_0$ and $H_1$ features were excluded, as they are not expected to capture (biologically relevant topological structures; exploratory analyses confirmed that their inclusion produced results almost identical to those obtained without them. The final feature set comprised 51 descriptors (see Appendix \ref{App:PCA_loadings} for their loadings on the first two PCA axes).

\begin{figure}[h!]
    \centering
    \begin{subfigure}[b]{0.46\textwidth}
        \centering
        \includegraphics[width=\textwidth]{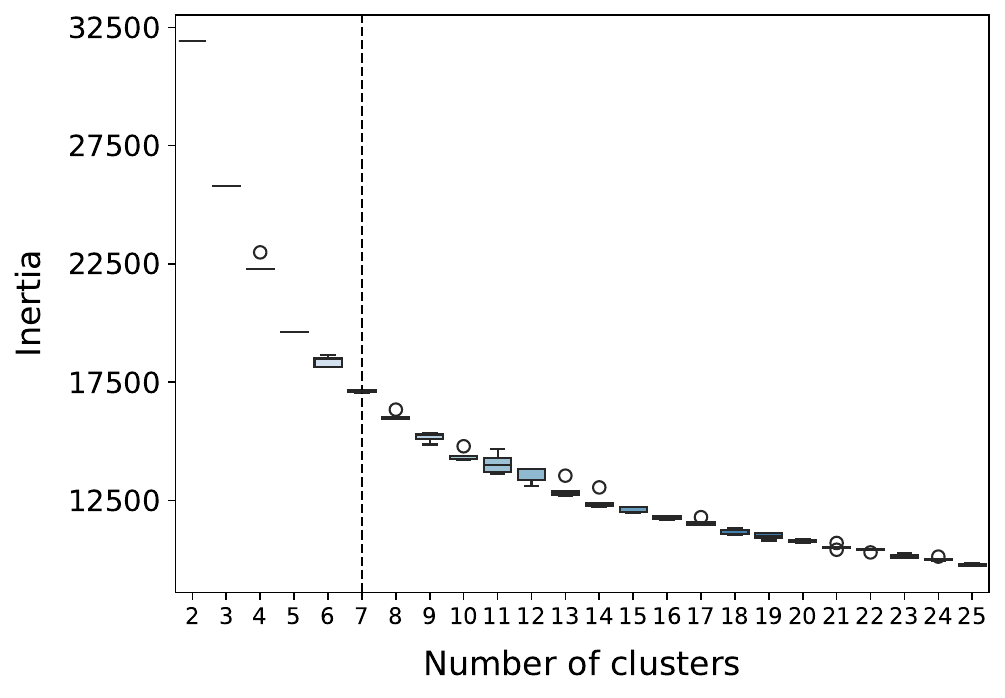} 
        \caption{}
        \label{subfig:Inertia}
    \end{subfigure}
    \hfill
    \begin{subfigure}[b]{0.46\textwidth}
        \centering
        \includegraphics[width=\textwidth]{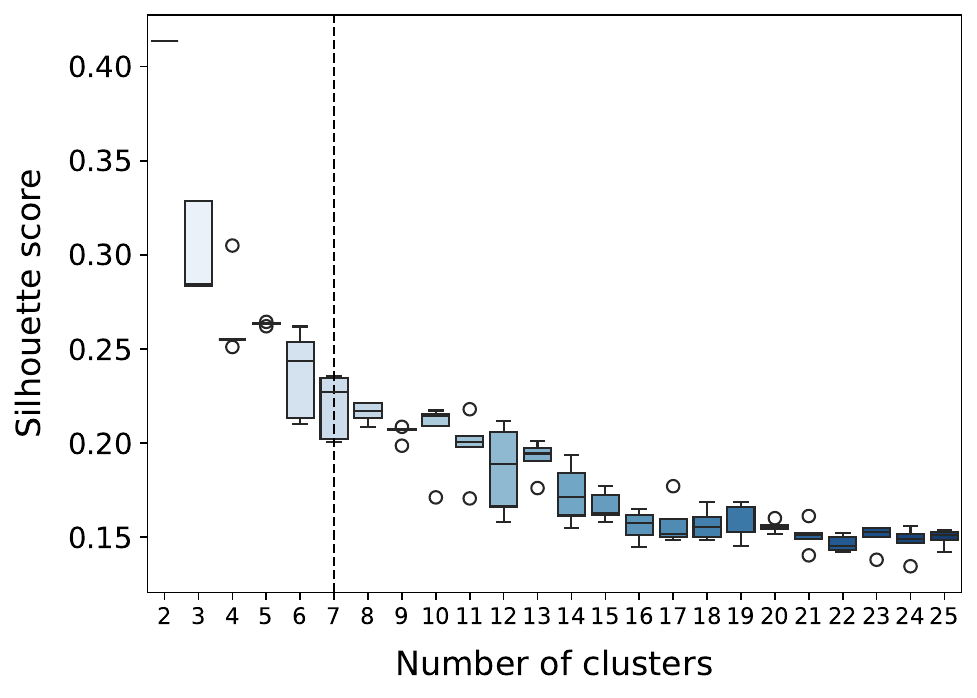} 
        \caption{}
        \label{subfig:Silhouette}
    \end{subfigure}

    \caption{Results of \textbf{(a)} inertia and \textbf{(b)} silhouette scores from 50 $k$-means runs for different numbers of clusters. The elbow at $k=7$ is indicated by a dashed line.}
    \label{fig:kmeansstats}
\end{figure}

The resulting feature vectors were z-score normalised prior to analysis. Principal component analysis (\texttt{scikit-learn}, \cite{scikit-learn}) was performed to assess correlation structure. For clustering, $k$-means (\texttt{scikit-learn}, \cite{scikit-learn}) was applied to the vectorised features. The optimal number of clusters ($k=7$) was selected by combining inertia and silhouette analysis (Figure \ref{fig:kmeansstats}). Comparable results were obtained for $k$ between 6 and 9. Median cluster centroids were computed, and a minimum spanning tree (MST) linking the centroids using correlation distance was constructed to visualise cluster relationships (see Appendix \ref{App:Centroid_coordinates} for evolution of centroid coordinates across the spanning tree).

Cluster enrichment by disease stage was tested using a hypergeometric test that compared, for each cluster, the observed and expected numbers of ROIs for each of the three pathologist-classified stages.

Hierarchical clustering of tissue samples based on their ROI cluster composition profiles was performed with \texttt{scipy.cluster.hierarchy} \cite{2020SciPy-NMeth}. Supplementary figures include full ROI partitions and cluster assignments for all datasets.

\textbf{Acknowledgements.} SSDHI was supported by Bowel Research UK (BRUK) grant number BKT00400. JWM was supported by Cancer Research UK (CRUK) grant number CTRQQR-2021\textbackslash100002, through the CRUK Oxford Centre. LG and SJL were supported by CRUK Program Grant (DRCNPG-Jun22\textbackslash100002) and the
MRC Mouse Genetics Network. EJM was supported by the Lee Placito Medical Research Fellowship (University of Oxford). SSDHI and HH are members of the Oxford/Max Planck collaboration and this research was funded in whole or in part by EPSRC international centre to centre collaboration grant  EP/Z531224/1. For the  purpose of Open Access, the authors have applied a CC BY public copyright licence to any Author Accepted Manuscript (AAM) version arising from this submission. SSDHI and HMB were supported by the CRUK Oxford Centre.

\bibliographystyle{vancouver}
\bibliography{references}

\clearpage
\appendix

\section{Plots of colorectal cancer dataset}\label{App:Real_datasets}

\begin{figure}[h!]
    \centering
    \includegraphics[width=0.85\textwidth]{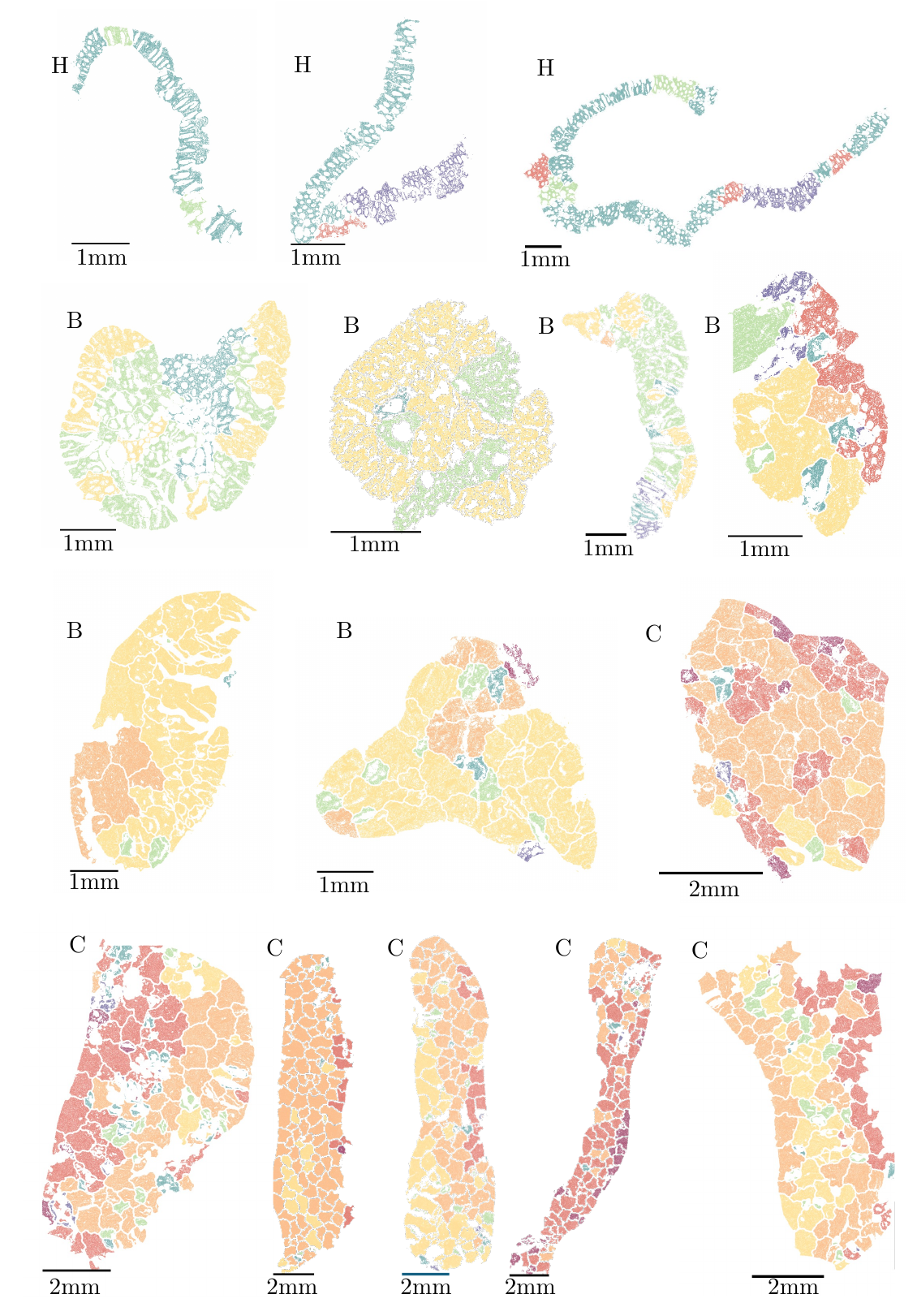} 
    
    \caption{Images of the 15 human colorectal tissue biopsies (3 healthy (H), 6 benign lesions (B), 6 cancerous (C)), each divided into regions of interest using TopROI. Regions are coloured according to the $k$-means clustering of their vectorised topological features (see Section \ref{subsec:Results_real}).}
    \label{fig:Datasets}
\end{figure}

\clearpage

\section{Resolution parameter determination}\label{App:Res_param_det}

To determine appropriate values of the resolution parameter $\gamma$, we applied a secant method under the assumption that the number of communities obtained increases monotonically with the resolution parameter. Pseudocode can be found in Algorithm \ref{alg:secant}. Since community detection is stochastic, exact convergence cannot always be guaranteed. If convergence is not achieved within a chosen number of iterations, we return the resolution parameter that yields the ROI count closest to the target value.

\begin{algorithm}[ht]
\caption{Secant method for determining the value of the resolution parameter $\gamma$}
\label{alg:secant}
\begin{algorithmic}[1]
\State \textbf{Input:} Target ROI number $N$, maximum iterations $K$, allowed error $\varepsilon$, precomputed list $L = \{(\gamma_i, n_i)\}$ of pairs (resolution parameter, number of ROIs)
\State $k \gets 0$
\While{no element in $L$ satisfies $\frac{|N - n_i|}{N} \leq \varepsilon$ \textbf{and} $k < K$}
    \State Identify $(\gamma_a, n_a), (\gamma_{a+1}, n_{a+1}) \in L$ with $n_a < N < n_{a+1}$
    \State $\gamma_{\text{new}} \gets \gamma_a + (N - n_a)\frac{\gamma_{a+1} - \gamma_a}{n_{a+1} - n_a}$
    \State Run community detection with $\gamma_{\text{new}}$ to obtain $n_{\text{new}}$
    \State Append $(\gamma_{\text{new}}, n_{\text{new}})$ to $L$
    \State $k \gets k + 1$
\EndWhile
\If{$k = K$}
    \State \textbf{Output:} Resolution parameter giving ROI number closest to $N$
\Else
    \State \textbf{Output:} Resolution parameter $\gamma_i$ with $\frac{|N - n_i|}{N} \leq \varepsilon$
\EndIf
\end{algorithmic}
\end{algorithm}

\begin{figure}[H]
    \centering
    \includegraphics[width=\textwidth]{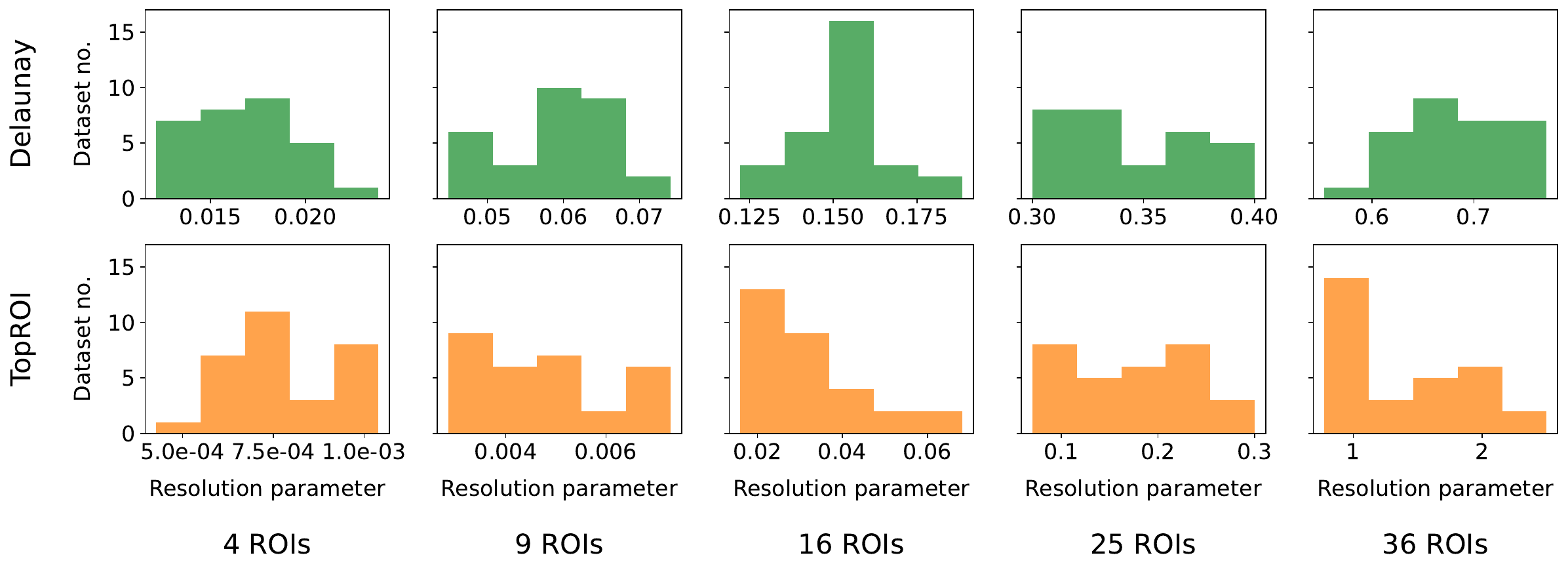} 
    
    \caption{Distribution of resolution parameter values for the synthetic dataset across methods and target number of ROIs.}
    \label{fig:synthetic_resolutions}
\end{figure}

We applied this approach to synthetic and biological datasets. For synthetic datasets, the initial list of values of the resolution parameters was $\{10^{-5}, 10^{-3}, 0.1, 0.5, 1, 2, 3, 4, 5\}$. From these starting points, the secant method was used to determine, for each dataset, the resolution parameter values that produced 4, 9, 16, 25, and 36 ROIs, with a tolerance of 0\% error. Across 300 runs (comprising different datasets, network construction methods -- Delaunay and TopROI -- and target ROI numbers), 17 runs (5.6\%) did not converge within 15 iterations. In these cases, the resulting ROI count differed from the target by one. The corresponding distributions of resolution parameter values are shown in Figure \ref{fig:synthetic_resolutions}.

For biological datasets, a single resolution parameter was required. We initialized the method with values of the resolution parameter of $10^{-5}$ and $1$, and accepted a maximum deviation of 5\% between the target and obtained ROI counts. All runs converged under this criterion. Table \ref{tab:real_resolutions} reports the resolution parameter values, target ROI numbers, obtained ROI numbers, and the number of iterations required for convergence.

\begin{table}[H]
\centering
\caption{Resolution parameter values for biological datasets, target/obtained ROI numbers, and iteration counts.}
\label{tab:real_resolutions}
\begin{adjustbox}{max width=\textwidth}
\begin{tabular}{cccccccccccccccc} 
&
\multicolumn{3}{c}{Healthy} & \multicolumn{6}{c}{Benign lesion} & \multicolumn{6}{c}{Cancer} \\
\cmidrule(lr){2-4} \cmidrule(lr){5-10} \cmidrule(lr){11-16}
Target ROI no.   & 9 & 15 & 40 & 17 & 20 & 33 & 45 & 47 & 53 & 76 & 120 & 150 & 151 & 153 & 160 \\
Obtained ROI no. & 9 & 15 & 40 & 17 & 20 & 34 & 46 & 48 & 53 & 74 & 122 & 157 & 150 & 159 & 164 \\
Resolution parameter& 0.0029 & 0.0086 & 0.020 & 0.0079 & 0.024 & 0.023 & 0.025 & 0.095 & 0.11 & 0.13 & 0.23 & 0.36 & 0.35 & 0.37 & 0.56 \\
Iterations          & 6 & 6 & 5 & 6 & 5 & 5 & 6 & 4 & 5 & 5 & 5 & 3 & 4 & 4 & 3 \\
\bottomrule
\end{tabular}
\end{adjustbox}
\end{table}

\section{Filtration values for cycle representatives}\label{App:Cycle_lifespan}

\begin{figure}[H]
    \centering
    \includegraphics[width=\textwidth]{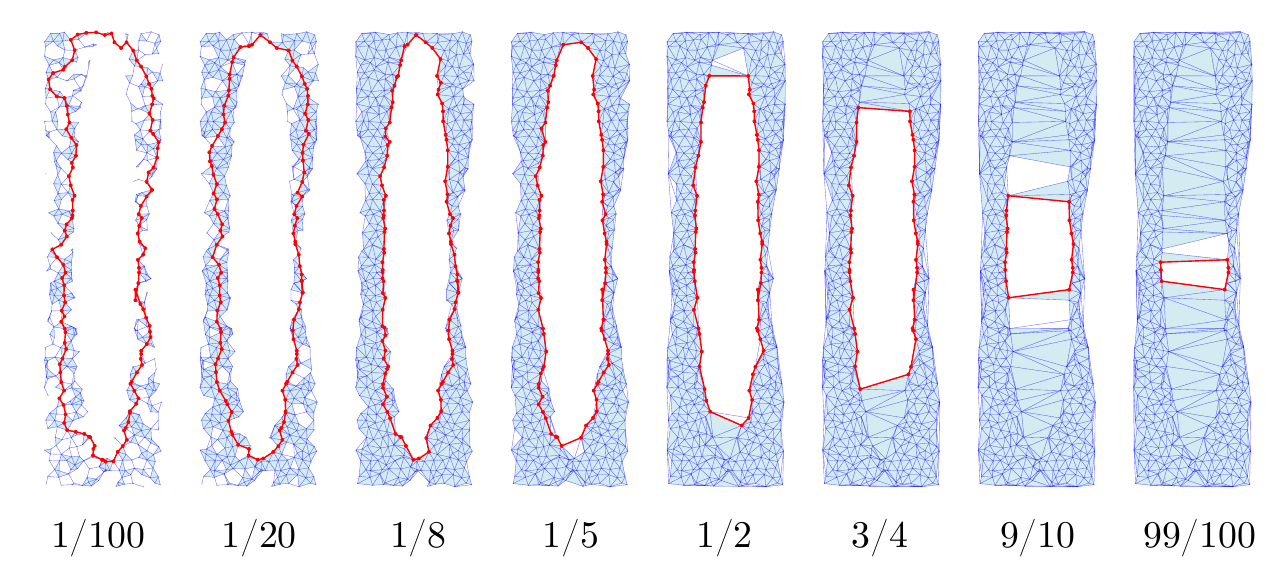} 
    
    \caption{Example of alpha complex (in blue) and computed cycle representative (red) at different time points of the cycle's lifespan. Picking an intermediate time yields a more geometrically faithful representative.}
    \label{fig:Cycles_filtration_values}
\end{figure}

\section{Elbow effect and resolution limit}\label{App:Resolution_limit}

\begin{figure}[h!]

    \centering
    \includegraphics[width=\textwidth]{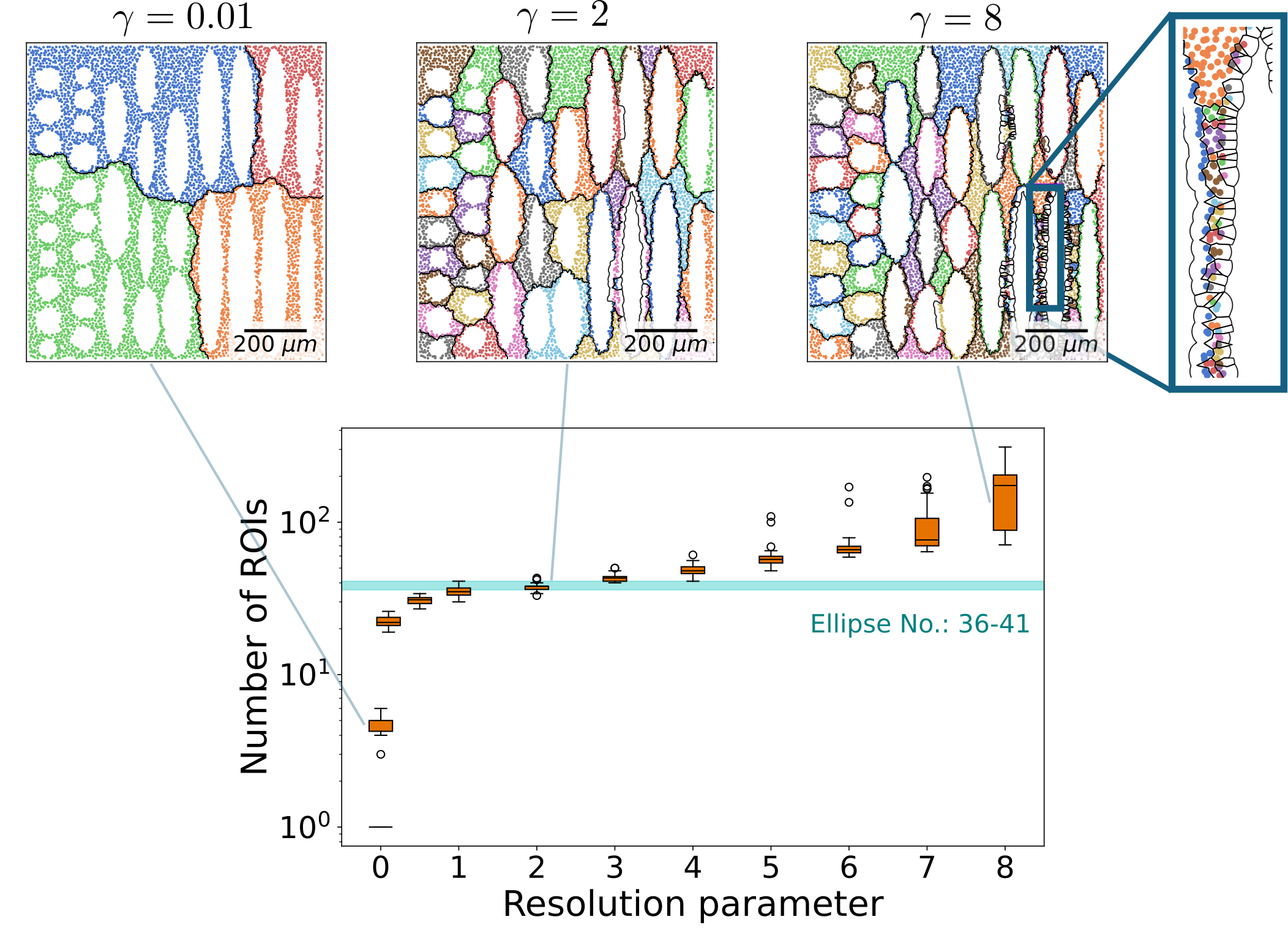} 
    \caption{Elbow effect and resolution limit in TopROI: dependence of the number of TopROI-generated ROIs on the resolution parameter $\gamma$ for the synthetic datasets. The elbow corresponds to the number of ellipses in the datasets. Example ROI partitions at selected values of $\gamma$ are also shown (black ROI borders are only for visualisation).}
    \label{fig:Elbow}
\end{figure}

We examined the effect of varying the resolution parameter, $\gamma$, in TopROI applied to the synthetic dataset. The number of ROIs increased sharply for $\gamma < 0.5$, then stabilised, with an elbow corresponding to the number of ellipses in the datasets (Figure \ref{fig:Elbow})). At low resolutions, TopROI communities captured topological features, with the Delaunay network extending partitions to adjacent points, until each topological feature was isolated within its own community.

At high values of the resolution parameter ($\gamma \geq 6$), point regions corresponding to the border of cycle representatives fragmented into single-point ROIs. This effect arises from the RBP quality function, and corresponds to the well-known resolution limit effect of modularity in community detection \cite{lancichinetti2011limits}]. For a complete graph with $n$ nodes and homogeneous edge weights $w$, the RBP score of the one-community partition is $n\cdot(n-1)\cdot w\cdot \left(1 - \gamma\right)$, whereas the score of the partition with each node in its own singleton community is $-(n-1)\cdot w \cdot \gamma$. Thus, for $\gamma > \frac{n-1}{n-2} \approx 1$, the singleton communities partition has a higher RBP score, and is favoured over the one-community partition. In the synthetic dataset, complete cycle graphs generated by TopROI are embedded within larger networks rather than appearing in isolation, which shifts this threshold to higher values of $\gamma$. Nevertheless, at sufficiently high values of $\gamma$, nodes belonging to cycle representatives tend to become isolated. This highlights the existence of a resolution threshold beyond which TopROI, in its current form, fails to produce meaningful communities. In practice, however, this limitation is not problematic, as all ROI numbers of interest in our experiments were achieved with resolution parameters below this threshold.

\section{Artefactual topological features}\label{App:Artefactual_figures}
\begin{figure}[H]
    \centering
    \begin{subfigure}[b]{0.64\textwidth}
        \centering
        \includegraphics[width=\textwidth]{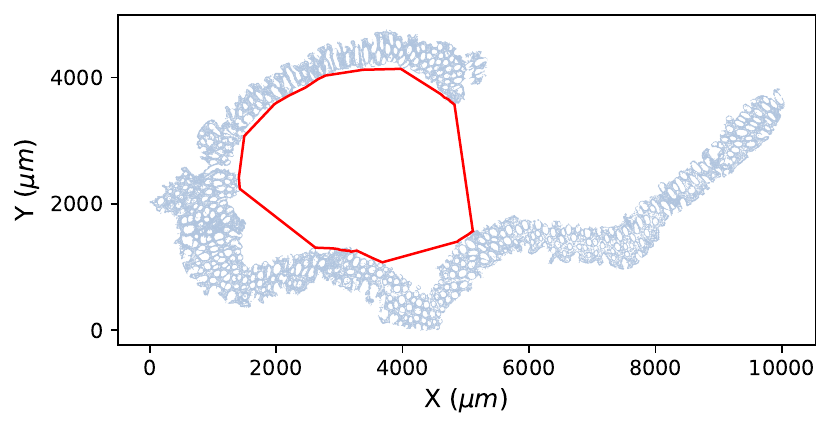} 
        \caption{}
        \label{subfig:Cycles_healthy}
    \end{subfigure}
    \begin{subfigure}[b]{0.35\textwidth}
        \centering
        \includegraphics[width=\textwidth]{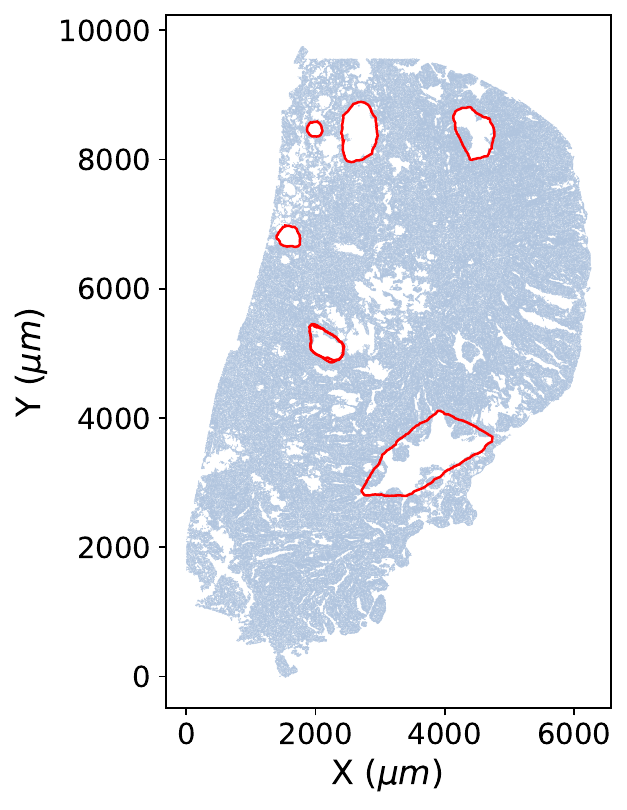} 
        \caption{}
        \label{subfig:Cycles_cancer}
    \end{subfigure}

    \caption{Examples of artefactual cycles with persistence greater than $100\mu m$ introduced during data preparation: \textbf{(a)} sample positioning artefacts, and \textbf{(b)} tissue tears produced during preparation.}
    \label{fig:Bad_cycles}
\end{figure}

\section{PCA loadings}\label{App:PCA_loadings}

\begin{figure}[H]
    \centering
    \includegraphics[width=\textwidth]{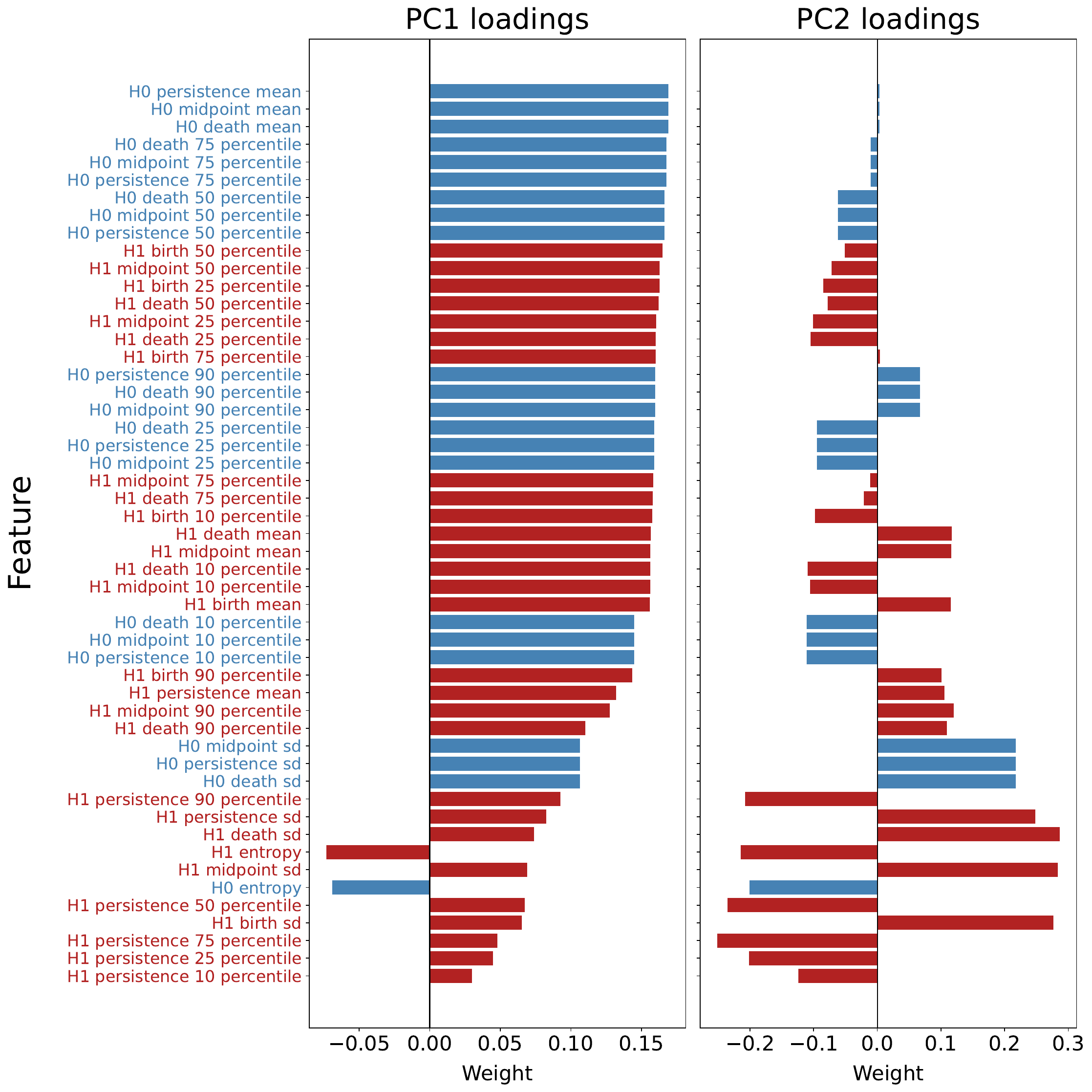} 

    \caption{Loadings of the 51 topological statistics of the first two principal components (see Section \ref{subsec:Results_real}). Bars indicate the contribution of each feature to the corresponding axis.}
    \label{fig:Loadings}
\end{figure}

\section{Cluster centroid coordinates}\label{App:Centroid_coordinates}

\begin{figure}[H]
    \centering
    \includegraphics[width=\textwidth]{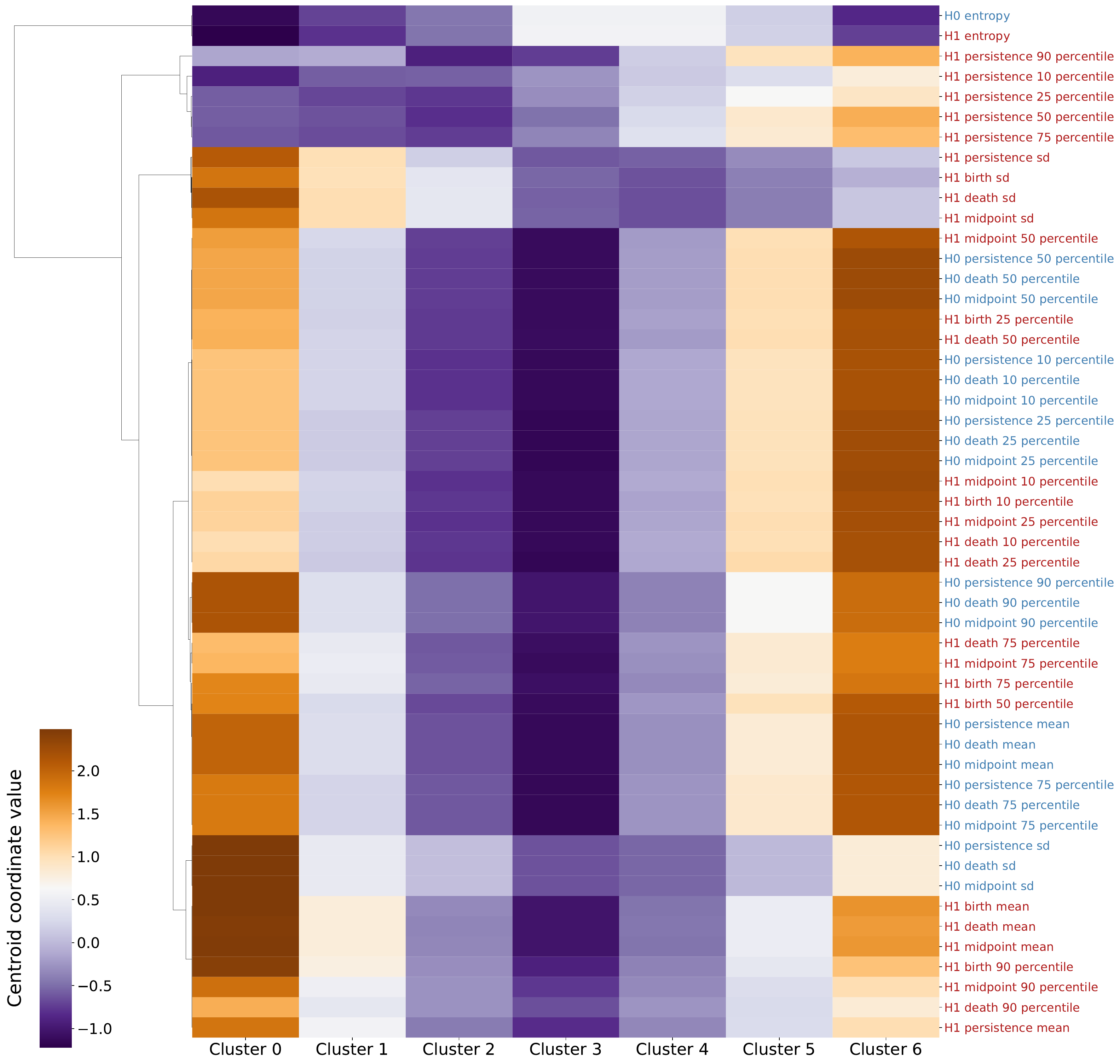} 

    \caption{Clustermap of the centroid coordinates for the seven $k$-means clusters (see Section \ref{subsec:Results_real}).}
    \label{fig:Clustermap}
\end{figure}

\end{document}

%% file: Preamble.tex
\usepackage[utf8]{inputenc}
\usepackage[english]{babel}
\usepackage{amssymb}
\usepackage{amsmath}
\usepackage{amsthm}
\usepackage{bigints}
\usepackage[labelfont=bf]{caption}
\usepackage{comment}
\usepackage{empheq}
\usepackage{textcomp}
\usepackage{enumerate}
\usepackage[shortlabels]{enumitem}
\usepackage{fancyhdr}
\usepackage{float}
\usepackage{geometry}
\usepackage{graphicx}
\usepackage{pdfpages}
\usepackage{lipsum}
\usepackage{listings}
\usepackage{mathtools}
\usepackage{multirow}
\usepackage{subcaption}
\usepackage{systeme}
\usepackage{tocbasic}
\usepackage{xfrac}
\usepackage[titles]{tocloft}
\usepackage{tikz}
\usetikzlibrary{matrix}
\usepackage{makecell}
\usepackage{nicefrac}
\usepackage{ulem}
\usepackage{lscape}
\usepackage{CJKutf8}
\usepackage{faktor}
\usepackage[numbers, sort&compress]{natbib}
\usepackage{booktabs}
\usepackage{adjustbox}
\usepackage{algorithm}
\usepackage{algpseudocode}
\usepackage{placeins}

\usepackage{hyperref}
\hypersetup{
	colorlinks=true,
	linkcolor=blue,
	citecolor=blue
}

\fancypagestyle{Appendix}{
	\fancyhead[L]{Appendix}
}
\fancypagestyle{References}{
	\fancyhead[L]{References}
}

\fancyheadoffset{1.5 cm} 
\fancypagestyle{plain}{%
	\fancyhf{}%
}



%% file: Commands.tex
\theoremstyle{definition}
